\documentclass[
reprint,
showpacs,
 amsmath,amssymb,
aps, pra,
showkeys,
]{revtex4-1}

\usepackage{graphicx}
\usepackage{dcolumn}
\usepackage{bm}

\begin{document}


\title{Resonant parametric interference effect in spontaneous bremsstrahlung of an
electron in the field of a nucleus and two pulsed laser waves}
\author{A.A. Lebed'}
\email{lebedo@ukr.net}
\author{E.A. Padusenko}%
 \email{lpadusenko@yandex.ru}
\affiliation{\\National Academy of Sciences of Ukraine, Institute of Applied
Physics, 58, Petropavlivska
St., 40000, Sumy, Ukraine }

\author{S.P. Roshchupkin}
\email{serg9rsp@gmail.com}
\author{V.V. Dubov} \email{dubov@spbstu.ru}
\affiliation{
Department of Theoretical Physics
Peter the Great St. Petersburg Polytechnic University,
29, Polytechnicheskaya St., 195251, Saint-Petersburg, Russia
}

\date{\today}

\begin{abstract}
Resonant spontaneous bremsstrahlung of an electron scattered by a nucleus in the field of two moderately
strong pulsed laser waves is studied theoretically. The process is studied in detail within the
interference kinematic region. This region is determined by scattering of particles in the same plane at
predetermined angles, at that stimulated absorption and emission of photons of external pulsed waves by an
electron occurs in correlated manner. The correspondence between the emission angle and the final-electron
energy is established in the kinematic region where the resonant parametric interference effect is
manifested. The resonant differential cross section of ENSB process with simultaneous registration of both
emission angles of the spontaneous photon and the scattered electron, can exceed by 4-5 orders of magnitude
the corresponding cross section in the absence of an external field.  It was shown for nonrelativistic
electrons that the resonant cross section of ENSB in the field of two pulsed laser waves within the
interference region in two order of magnitude may exceed corresponding cross section in the Bunkin-Fedorov
kinematic region. The obtained results may be experimentally verified, for example, by scientific
facilities at sources of pulsed laser radiation (SLAC, FAIR, XFEL, ELI).
\end{abstract}


\maketitle


\section{Introduction}

Nonlinear effects of quantum electrodynamics (QED) in force fields have been an object of scientific
research for a long time already. At the same time, studying of these effects is still relevant in our
time. This interest is due to continuous development of laser technology and experimental systems for
testing QED effects \cite{mourou}. Improvement of laser systems, generally, consists in production of
increasingly short and intense laser pulses. New experimental conditions have required constant
improvements in calculations and model development \cite{Piazza_rev}.

Electron motion becomes highly nonlinear as a function of the laser electromagnetic field, under the
relativistic regime, in laser-electron interaction. The relativistic-regime threshold (the laser intensity
is higher than $10^{18}\,\rm{W\!\cdot \!cm}^{-2}$) has been already reached, and even exceeded in the
world's leading scientific laboratories. Thus, laser systems which can provide field powers of the order of
1 PW were constructed (Vulcan, Vulcan10, PHELIX, XFEL ) \cite{Piazza_rev,Phelix}. Experimental verification
of QED effects in the laser field was carried out at the facility SLAC National Accelerator Laboratory
(Stanford, USA) \cite{Bula,Burke}, and also is included into scientific programs of FAIR (Facility for
Antiproton and Ion Research), international project at the GSI in Darmstadt (Germany) based on the laser
system PHELIX \cite{Phelix}, and also into the programs of the Extreme Light Infrastructure (ELI) project.
The scientific interest to studying of QED processes in laser fields of such intensities is caused by the
possibility of testing of different aspects of fundamental physics for the first time.

Presence of an external laser field in QED processes can considerably affect the magnitude of the cross
section, the angular distribution and energy spectra of the final particles. QED processes of both the
first and second order in the field of a laser wave were studied in Refs.
\cite{Bula,Burke,laes,Nar_fof,Bunkin_Fedorov,Fedorov,monmon,monpul,intech,
monmon,monpul,intech,Ritus,Ehlotzky_rev,Ehlotzky_rev2,Nonres_rev,Res_rev_rsp,Res_rev,
karap_fed,Zhou93,krainov_rsp_slow,lebedev72,borisovSBRES,rspNucFiz,rsp_2002,Dondera,
Florescu1,zhelt,Flegel,zhelt2,Li,LotstedtSB,schnezSB,Nonres_sb,Leb_sb_NN,
LebResSB,Rsp_lys_two,Rsp_lys_two2,Gorod,Rsp_1994,Rsp_1996,rsp_leb_en_two,pie_sb,pie_cpp,
Rsp_vor_two,Rsp_vor_two2,Krajewska_2012_comton}. The results have been summarized in monographs
\cite{Fedorov,monmon,monpul,intech} and reviews
\cite{Ritus,Ehlotzky_rev,Ehlotzky_rev2,Nonres_rev,Res_rev_rsp,Res_rev}.

QED nonlinear effects in force fields are closely related to the processes kinematics. The second-order
processes can possibly proceed in the resonant manner in the specified kinematics. This particularity is
well-known. The resonant character is due to the fact that certain laser-induced processes of the first
order are allowed in the laser field, but they do not occur in the absence of an external field. The
particle in the intermediate state can fall within the mass shell in a certain range of values of the
energy and momentum. Consequently, the considered higher-order process effectively reduces into two
successive lower-order processes. Wherein, values of resonant cross sections can exceed values of
corresponding cross sections in the absence of an external field, by several orders of the magnitude
\cite{Res_rev_rsp,Res_rev}. The appearance of resonances in the laser field is among the fundamental
problems of QED in the strong fields.

An electron, when it is scattered by a nucleus in the external laser field, can forcedly emit and absorb
external-field photons and spontaneously radiate a photon of the arbitrary frequency. This is
laser-modified electron-nucleus spontaneous bremsstrahlung (ENSB) process. Bremsstrahlung is one of the
main mechanisms of energy loss by an electron, when interacting with the substance. Spontaneous
bremsstrahlung (SB) of an electron scattered by an atom or a nucleus in the external electromagnetic field
is of interest for quite a long time \cite{monmon,monpul,intech,Nonres_rev,Res_rev_rsp,Res_rev,
karap_fed,Zhou93,krainov_rsp_slow,lebedev72,borisovSBRES,rspNucFiz,
rsp_2002,Dondera,Florescu1,zhelt,Flegel,zhelt2,Li,LotstedtSB, schnezSB,Nonres_sb,Leb_sb_NN,
LebResSB,Rsp_lys_two,Rsp_lys_two2,pie_sb}. This process under resonance conditions in the plane
electromagnetic field was considered in works \cite{lebedev72,borisovSBRES,rspNucFiz,LotstedtSB,LebResSB}.
Resonant ENSB for electron nonrelativistic energy in the plane-wave field was studied by Lebedev
\cite{lebedev72}. Borisov et al. \cite{borisovSBRES} considered resonant SB that accompanies collisions of
ultrarelativistic electrons for large transferred momenta. In the general relativistic case, the problem of
ENSB in the field of a plane monochromatic wave was studied by Roshchupkin
\cite{Res_rev_rsp,rspNucFiz,rsp_2002,monmon}. Resonant case of ENSB in the pulsed laser field was
considered in the Ref. \cite{LebResSB}. It was manifested that consideration of the external-field pulsed
character eliminates the resonant divergence in the process cross section. The magnitude of the resonant
cross-section is strongly dependent on the wave length and electron initial energy.

When considering QED processes in the external field, one can distinguish the case when the laser field is
a superposition of two plane waves
\cite{Rsp_lys_two,Rsp_lys_two2,Gorod,Rsp_1994,Rsp_1996,rsp_leb_en_two,pie_sb,pie_cpp,
Rsp_vor_two,Rsp_vor_two2,Krajewska_2012_comton}. Herein, the cross section has the form of the sum of
partial components. Each of components corresponds to emission and absorption of a certain number of
photons of the first and second waves. The parametric interference effect manifests when QED processes
proceed in the field of two laser waves. The essence of this effect is that within the specified kinematics
(the interference region) scattering particles can forcedly absorb and emit photons of electromagnetic
waves in correlated manner \cite{Gorod,Rsp_1994,Rsp_1996}. Moreover, in the case of circular polarization,
processes with emission and absorption of the equal number of photons of the first and second laser wave
can predominate. Study of ENSB in the field of two laser waves was carried out for the nonresonant case and
plane monochromatic waves in Refs. \cite{Rsp_lys_two,Rsp_lys_two2}. Nonresonant ENSB in the field of two
moderately strong pulsed laser waves was studied in Ref. \cite{pie_sb}. It was concluded that the
probability of the partial process with correlated emission (absorption) by an electron of the equal number
of photons of both waves is of an order of the magnitude greater than the corresponding probability in any
other scattering kinematics.

It is of interest to study in detail the resonant parametric interference effect, i.e. to study the process
kinematics for simultaneous implementation of resonance conditions and correlation between emission from
the first and second wave. It should be noted that resonance and interference effects are of different
nature, and considerably affect the value of the differential cross section of processes. In the present
work we develop a theory of resonant SB of an electron scattered by a Coulomb center in presence of an
external field of two pulsed electromagnetic waves. The main aim of the work is detailed analysis of
resonances of the studied process within the interference region, where peculiar properties of stimulated
absorption and emission of photons of waves by an electron appear.

\subsection{External laser field} The external pulsed field was chosen as a superposition of two plane
non-monochromatic waves, propagating in the same direction along the $z$ axis, with the plane of
polarization ($xy$). The four-potential of such a field has the form
\begin{equation} \label{ZEqnNum318473}
A\left(\varphi _{1} ,\varphi _{2} \right)={\it g}_{1}\! \left(\frac{\varphi _{1} }{\omega _{1} \tau _{1} }
\right)A_{1\left({\rm mon}\right)} +{\it g}_{2}\! \left(\frac{\varphi _{2} }{\omega _{2} \tau _{2} }
\right)A_{2\left({\rm mon}\right)} ,
\end{equation}
\begin{equation} \label{ZEqnNum393125}
A_{j\left({\rm mon}\right)} =\frac{cF_{0j} }{\omega _{j} } \left(e_{jx} \cos \varphi _{j} {+}\delta _{j}
e_{jy} \sin \varphi _{j} \right),\quad \!\!j=1,2,
\end{equation}
\begin{equation} \label{ZEqnNum195147}
\varphi _{j}  =\left(k_{j} x\right)=\omega _{j} \xi ,\quad \xi =t-z/c.
\end{equation}
Each of summands in the Eq. \eqref{ZEqnNum318473} corresponds to the field of the first and second pulsed
laser wave (the index $j$ labels the wave) and $\varphi _{j} $ is the wave phase and $\tau _{j} $ is the
pulse width. In the Eqs. \eqref{ZEqnNum318473}--\eqref{ZEqnNum195147} $c$ is the light velocity in vacuum;
$F_{0j} $ is the strength amplitude of the electric field in the pulse peak; $\omega _{j} $ is the
laser-wave characteristic frequency; $k_{j} =\left(\omega _{j}/c ,{\bf k}_{j} \right)$ is the wave
four-vector; $\delta _{j} $ is the wave ellipticity parameter ($\delta _{j} =0$ corresponds to linear
polarization, $\delta _{j} =\pm 1$ corresponds to circular polarization); and $e_{jx} =\left(0,{\bf e}_{jx}
\right)$ and $e_{jy} =\left(0,{\bf e}_{jy} \right)$ are four-vectors of wave polarization, meeting the
conditions:
\begin{equation} \label{4}
e_{jx,jy}^{2} =-1, \quad \left(e_{jx,jy} k_{j} \right)=k_{j}^{2} =0.
\end{equation}
Hereafter, the standard metric for four-vectors, $\left(ab\right)=a_{0} b_{0} -{\bf ab}$, is used.

Functions $ g_{j} \left({\varphi _{j} / \omega _{j} \tau _{j} } \right)$ in Eq. \eqref{ZEqnNum318473} are
envelope functions of the four-potential of an external wave, that allows to take into account the pulsed
character of a laser field \cite{Nar_fof}. The process is studied within the frame of the
quasimonochromatic approximation, when a laser wave performs a lot of amplitude oscillation, i.e. the
following condition is met:
\begin{equation} \label{ZEqnNum305876}
\omega _{j} \tau _{j} \gg 1.
\end{equation}
The condition \eqref{ZEqnNum305876} is satisfied for the majority of modern lasers
\cite{mourou,Piazza_rev}. We emphasize that an electromagnetic field with the four-potential
\eqref{ZEqnNum318473}--\eqref{ZEqnNum195147} represents a plane wave. Thereby Volkov functions
\cite{Volkov,Landau}, which are correct for a plane wave of arbitrary spectral composition, can be used for
description of the electron state in the field of a quasimonochromatic wave.

Note that in description of QED processes in the laser field it is convenient to use the classical
relativistic-invariant multiphoton parameter \cite{Ritus}
\begin{equation} \label{ZEqnNum897549}
\eta _{0j} =\frac{eF_{0j} }{mc\omega _{j} } ,
\end{equation}
where, $e$ is an electron charge, $m$ is an electron mass. It numerically equals to the ratio of the work
done by the field over an electron, on the wavelength, to the electron rest energy. In the classical
consideration of laser-dressed electron motion the parameter $\eta_{0j}$ defines the characteristic
velocity of electron oscillation in the case if $\eta _{0j} \ll 1$. The problem of ENSB will be studied in
the range of moderately strong fields, when
\begin{equation} \label{ZEqnNum786952}
\eta _{0j} \ll 1.
\end{equation}
Conditions (\ref{ZEqnNum786952}) meets the typical range of field strengths for modern laser facilities:
$F_{0j}  \ll 10^{10} \div 10^{11} \,\rm{V/cm} $.

In what follows we consider the case of circular polarization of external pulsed waves:
\begin{equation} \label{ZEqnNum471197}
\delta _{1} =+1, \quad \delta _{2} =\mp 1.
\end{equation}
The case $\delta _{1} =\delta _{2} $ corresponds to rotation of vectors of the field strength in the same
direction; in the opposite, the case $\delta _{1} =-\delta _{2} $ corresponds to rotation in the opposite
directions.

It should be noted that in the case of waves' close frequency and same polarization, we have the case of a
single wave \cite{Rsp_1994}. Note also that description of a laser field by the potential
\eqref{ZEqnNum318473}--\eqref{ZEqnNum195147} does not take into account the possible phase shift between
laser waves and stipulates that laser pulses' maxima coincide.

The relativistic system of units, $\hbar =c=1$, will be used throughout this paper.

\section{Amplitude of ENSB process in two laser waves}

Let us consider emission of a photon with the wave four-vector $k'{=}\left(\omega ',{\bf k}'\right)$ if an
electron in the state with the four-momentum $p_{i}{=}\left(E_{i} ,{\bf p} _{i} \right)$ is scattered by a
nucleus into the state with the four-momentum $p_{f} {=}\left(E_{f} , {\bf p}_{f} \right)$ in the field of
two pulsed laser waves.

\begin{figure}
\includegraphics[width=8.5cm]{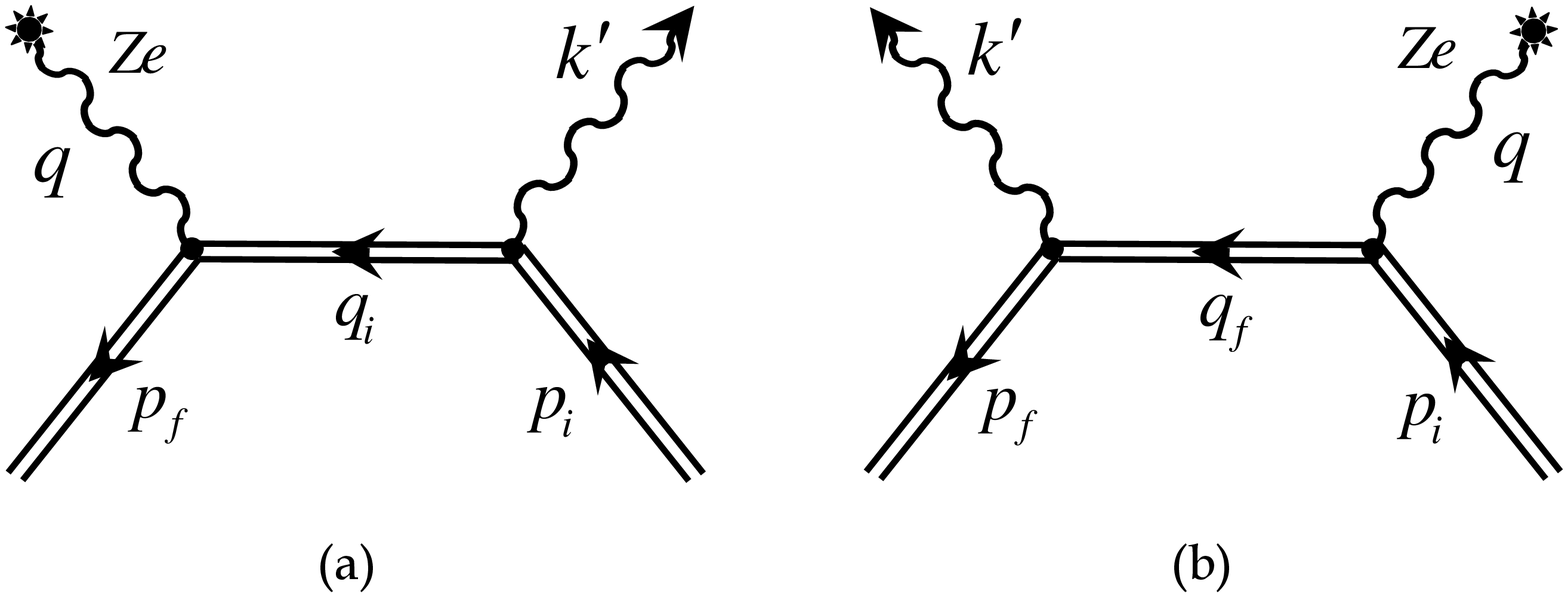}
\caption{\label{figure1} Feynman diagrams of spontaneous bremsstrahlung of an electron scattered by a
nucleus in the field of two pulsed waves. Dual incoming and outgoing lines correspond to wave functions of
an electron in the initial and final states, the inner line corresponds to the Green function of an
electron in the field of two pulsed waves. Wavy lines correspond to four-momenta of the spontaneous photon
and the photon of recoil of a nucleus.}
\end{figure}

Electron interaction with a nucleus is considered in the frame of the Born approximation:
\begin{equation} \label{9)}
v_{i,f} \gg Z\alpha ,
\end{equation}
where $v_{i,f} ={\left|{\bf p}_{i,f}\right|/ E_{i,f} } $ is the electron velocity before and after
scattering; $Z$ is the nucleus charge number and $\alpha $ is the fine-structure constant. ENSB process is
described by two Feynman diagrams within the frame of the Born approximation (see, Fig. \ref{figure1}).

The amplitude of the considered process in the field of two moderately strong pulsed electromagnetic waves
\eqref{ZEqnNum318473}-\eqref{ZEqnNum195147}, in the general relativistic case  may be represented in the
form \cite{Rsp_lys_two,pie_sb}
\begin{equation} \label{ZEqnNum750977}
S_{fi} =\sum _{l_{1} ,l_{2} =-\infty }^{\infty }S_{l_{1} l_{2} }.
\end{equation}
Here $S_{l_{1} l_{2} } $  is the partial amplitude of ENSB with absorption ($l_{1} ,l_{2} <0$) or emission
($l_{1} ,l_{2} >0$) of photons of external laser field:
\begin{equation} \label{ZEqnNum101734}
S_{l_{1} l_{2} } =-i\frac{Ze^{3} \sqrt{\pi } }{\sqrt{2\omega 'E_{i} E_{f} } } \,\bar{u}_{f}
\!\left[B_{l_{1}\! l_{2} }^{\left(a\right)}\! \left(q_{i} \right)+B_{l_{1}\! l_{2} }^{\left(b\right)}
\!\left(q_{f} \right)\right]\!u_{i} ,
\end{equation}
where, $u_{i} $ and $\bar{u}_{f} $ are Dirac's bispinors functions; indices $l_{1} $ and $l_{2} $ are the
number of emitted (absorbed) photons for the first and second wave, respectively. Functions $B_{l_{1} l_{2}
}^{\left(a\right)} \left(q_{i} \right)$ and $B_{l_{1} l_{2} }^{\left(b\right)} \left(q_{f} \right)$
correspond to Figs. \ref{figure1}(a) and \ref{figure1}(b) of the considered process, respectively:
\begin{eqnarray} \label{ZEqnNum122312}
&&\displaystyle{B_{l_{1}\! l_{2} }^{\left(a\right)} \left(q_{i} \right)=2\omega _{1} \!\!\sum _{s_{1}
,s_{2}
=-\infty }^{\infty }\int _{-\infty }^{\infty }\!d\zeta }\\
&&\displaystyle{\times\frac{\tilde{M}_{l_{1} {+}s_{1}\!,\, l_{2} {+}s_{2} }^{0}\! \left(p_{f} ,q_{i} ,\zeta
\right)\left[\hat{q}_{i} +m\right]\!\left(\varepsilon _{\mu } ^{*} \tilde{M}_{{-}s_{1} ,-s_{2} }^{\mu }
\!\left(q_{i} ,p_{i} ,\zeta \right)\!\right)}{\left[q_{i}^{2} -m^{2} +2\zeta \left(k_{1} q_{i}
\right)+i0\right]\left[q_{\bot }^{2} +\left(q_{0} -q_{z} \right)^{2} \right]} \nonumber}   ,
\end{eqnarray}
\begin{eqnarray} \label{ZEqnNum108479}
&&\displaystyle{B_{l_{1}\! l_{2} }^{\left(b\right)} \left(q_{f} \right)=2\omega _{1} \!\!\sum _{s_{1}
,s_{2}
=-\infty }^{\infty}\int _{-\infty }^{\infty }\!d\zeta }\\
&&\displaystyle{ \times \frac{\left(\varepsilon _{\mu } ^{*} \tilde{M}_{-s_{1} ,-s_{2} }^{\mu }\!
\left(p_{f} ,q_{f} ,\zeta \right)\!\right)\!\left[\hat{q}_{f} \!+m\right]\tilde{M}_{l_{1}
{+}s_{1}\!,\,l_{2} {+}s_{2} }^{0}\! \left(q_{f} ,p_{i} ,\zeta \right)}{\left[q_{f}^{2} -m^{2} +2\zeta
\left(k_{1} q_{f} \right)+i0\right]\left[q_{\bot }^{2} +\left(q_{0} -q_{z} \right)^{2} \right]} \nonumber}.
\end{eqnarray}
The four-vector $q=\left(q_{0} ,{\bf q}\right)$ makes sense of the transferred four-momentum to the
nucleus; $q_{i} $ is the four-momentum of an electron in the intermediate state for the  diagram (a),
$q_{f} $ is the four-momentum of an electron in the intermediate state for the diagram (b) in
Fig.~\ref{figure1}:
\begin{eqnarray} \label{ZEqnNum249079}
&&{q_{i} =p_{i} -k'+s_{1} k_{1} +s_{2} k_{2} ,} \\ &&{q_{f} =p_{f} +k'+\left(l_{1} +s_{1} \right)k_{1}
+\left(l_{2} +s_{2} \right)k_{2} ,} \\ &&{q=p_{f} -p_{i} +k'+l_{1} k_{1} +l_{2} k_{2} .}
\end{eqnarray}
The transferred four-momentum $q$ is determined by the sum $l_{1} k_{1} +l_{2} k_{2} $, this sum specifies
the number of external-field photons, forcedly absorbed or emitted by an electron in the ENSB process.
Indices $s_{1} $ and $s_{2} $ are the number of photons in virtual emission ($s_{1,2} <0$) or absorption
($s_{1,2} >0$) for the first and second wave, respectively. They cannot be directly measured in the general
case of the process kinematics.

Let us introduce new dimensionless integration variables $\phi _{r} $:
\begin{eqnarray} \label{ZEqnNum423530}
&&\displaystyle{\phi _{r} =\frac{\xi _{r} }{\tau _{1} } ,\quad r=1,2;
}\\
&&\displaystyle{ \xi _{1} =\left(nx_{1} \right)=t_{1} -z_{1} ,\quad \xi _{2} =\left(nx_{2} \right)=t_{2}
-z_{2},\nonumber}
\end{eqnarray}
where $n\equiv \left(1,{\bf n}\right)=k_{j} /\omega _{j}$, ${\bf n}$ is a unit vector along the direction
of propagation of laser waves; $r$ index labels the vertex in Feynman diagrams.

In Eqs. \eqref{ZEqnNum122312} and \eqref{ZEqnNum108479} the integral function $M_{l_{1} {+}s_{1} ,l_{2}
{+}s_{2} }^{0} $ corresponds to electron-nucleus scattering \cite{laes,rsp_leb_en_two}, and the function
$M_{s_{1} s_{2} }^{\mu } $ corresponds to emission of a spontaneous photon \cite{Rsp_vor_two,Rsp_vor_two2}.
They are determined by integrals over wave variables \eqref{ZEqnNum423530} as:
\begin{equation} \label{ZEqnNum504293}
\tilde{M}_{n_{1}\! n_{2} }^{0} \!\left(p,p'\!,\zeta \right)=\tau _{1} \!\int \!d\phi _{1}\, {\rm e}^
{i\left(q_{i0} {-}\zeta \omega _{1} \right)\tau _{1} \phi _{1} } M_{n_{1}\! n_{2} }^{0}\!
\left(p,p'\right)\! ,
\end{equation}
\begin{equation} \label{ZEqnNum947972}
\tilde{M}_{n_{1}\! n_{2} }^{\mu } \left(p,p'\!,\zeta \right)=\tau _{1} \!\int \!d\phi _{2} \, {\rm e}^
{i\zeta \omega _{1} \tau _{1} \phi _{2} } M_{n_{1} n_{2} }^{\mu }\! \left(p,p'\right) \!,
\end{equation}
\begin{eqnarray}\label{ZEqnNum233516}
&&\displaystyle{M_{n_{1}\! n_{2} }^{\mu } \!\left(p,p'\right)=I_{n_{1}\! n_{2} } \!\left(\phi _{r}
\right)\tilde{\gamma }^{\mu } +\frac{m^{2} }{2\!\left(np'\right)\!\left(np\right)} B_{n_{1}\! n_{2} }
\!\left(\phi _{r} \right)n^{\mu } \hat{n}
\nonumber}\\
&&\displaystyle{ +\frac{m}{4\!\left(np'\right)} \hat{D}_{n_{1}\! n_{2} }\! \left(\phi _{r}
\right)\hat{n}\tilde{\gamma }^{\mu } +\frac{m}{4\!\left(np\right)}\tilde{\gamma }^{\mu } \hat{n}
\hat{D}_{n_{1}\! n_{2} }\! \left(\phi _{r} \right),}
\end{eqnarray}

The physical meaning of the parameter $\zeta $ is the energy spread of an electron in an intermediate state
in ENSB in the field of two pulsed laser waves in units of the energy of the first-wave photon. The
quasimonochromatic condition \eqref{ZEqnNum305876} results in sharp narrowing of the essential range of
integration variable in Eqs.~\eqref{ZEqnNum122312} and \eqref{ZEqnNum108479}. It is determined by the
following condition:
\begin{equation} \label{ZEqnNum537495}
\zeta \lesssim \left(\omega _{j} \tau _{j} \right)^{-1} \ll 1.
\end{equation}
As a result of accounting of a pulsed character of the field, the denominator in Eqs.~\eqref{ZEqnNum122312}
and \eqref{ZEqnNum108479} contains corrections, which depend on the variable $\zeta $. The similar
correction is absent in the monochromatic-wave case. It results in the resonant infinity in the amplitude
of ENSB process in the field of a plane monochromatic wave.

Functions $B_{n_{1}\! n_{2} }\! \left(\phi _{r} \right)$ and $D_{n_{1}\! n_{2} } \!\left(\phi _{r} \right)$
in Eqs. \eqref{ZEqnNum504293} and \eqref{ZEqnNum947972} have the form in the case of circular polarization
\eqref{ZEqnNum471197}:
\begin{eqnarray}\label{ZEqnNum298670}
&&\displaystyle{B_{n_{1} \!n_{2} } \!\left(\phi _{r} \right)\!=\! \left(\eta _{1}^{2}\! \left(\phi _{r}
\right)+\eta _{2}^{2}\! \left(\phi _{r} \right)\right) I_{n_{1}\! n_{2} } \!\left(\phi _{r} \right)
}\\
&&\displaystyle{+\eta _{1}\! \left(\phi _{r} \right) \eta _{2}\! \left(\phi _{r} \right)\!\left[{\rm
e}^{-i\Delta } \!I_{n_{1} {+}1\!,n_{2} {\pm} 1}\! \left(\phi _{r} \right){+}{\rm e}^{i\Delta }\! I_{n_{1}
{-}1\!,n_{2} {\mp} 1}\! \left(\phi _{r}
\right)\!\right]\!, \nonumber}\\
\label{ZEqnNum159541}
 &&\displaystyle{D_{n_{1}\! n_{2} }\! \left(\phi _{r} \right){=}\eta _{1}\! \left(\phi _{r} \right)
\!\left[\varepsilon _{1}^{\left(+\right)}\! I_{n_{1} -1,n_{2} }\! \left(\phi _{r} \right) {+}\varepsilon
_{1}^{\left(-\right)}\! I_{n_{1} +1,n_{2} } \!\left(\phi _{r} \right)\right]\nonumber}\\
&&\displaystyle{+\eta _{2} \!\left(\phi _{r} \right)\! \left[\varepsilon _{2}^{\left(+\right)}\! I_{n_{1}
,n_{2} -1}\! \left(\phi _{r} \right)+\varepsilon _{2}^{\left(-\right)} \!I_{n_{1} ,n_{2} +1}\! \left(\phi
_{r} \right)\right],}\\ \label{ZEqnNum904263}
 &&\displaystyle{\eta _{j}\! \left(\phi _{r} \right)=\eta _{01} g_{j} \!\left(\phi _{r} \right),\quad\varepsilon _{j}^{\left(\pm
\right)} =e_{jx} \pm i\delta _{j} e_{jy} .}
\end{eqnarray}
The special functions $I_{n_{1} n_{2} } \left(\phi _{r} \right)$ in Eqs. \eqref{ZEqnNum298670} and
\eqref{ZEqnNum159541} determine the probability of partial multiphoton processes in the field of two pulsed
laser waves. These functions are studied in detail in Refs. \cite{L_fun}. They can be represented in the
form of expansion into series of Bessel functions. For circular polarization we obtain:
\begin{eqnarray}\label{ZEqnNum482342}
&&\displaystyle{I_{n_{1}\! n_{2} } \left(\chi _{r} ,\gamma _{r} ,\alpha _{\pm } \right)=\exp
\left\{-i\left(n_{1} \chi _{1}^{} +n_{2} \chi _{2}^{} \right)\right\}
\nonumber}\\
&&\displaystyle{\times\!\!\sum _{s=-\infty }^{\infty }\!\!{\rm e}^{is\left(\chi _{1} \pm \chi _{2} -\Delta
\right)} J_{s}\! \left(\alpha _{\pm } \right)J_{n_{1} -s} \!\left(\gamma _{1} \right) J_{n_{2} \mp s}
\!\left(\gamma _{2} \right) \!.\quad }
\end{eqnarray}
The sign $"\pm "$ in Eqs. \eqref{ZEqnNum298670} and \eqref{ZEqnNum482342} corresponds to the selected
direction of rotation of vectors of field strength \eqref{ZEqnNum471197}. Note that when the arguments of
these functions are independent from the indices, we have:
\begin{equation} \label{ZEqnNum662245}
\sum _{n_{1} ,n_{2} =-\infty }^{\infty }\left|I_{n_{1}\! n_{2} } \left(\chi _{r} ,\gamma _{r} ,\alpha _{\pm
} \right)\right|^{2}  =1 .
\end{equation}
Arguments of functions \eqref{ZEqnNum482342}, in general relativistic case and for circular polarization,
are determined as:
\begin{eqnarray}\label{ZEqnNum684936}
&&\displaystyle{\gamma _{j}\! \left(p,p',\phi _{r} \right)=\gamma _{0j}\! \left(p,p'\right)\cdot g_{j}\!
\left(\phi _{r} \right),
}\\
&&\displaystyle{\label{ZEqnNum680578} \gamma _{0j}\! \left(p,p'\right)=\eta _{0j} \frac{m}{\omega _{j} }
\sqrt{-Q_{\!pp'}^{2} } , \quad \! Q_{\!pp'} =\frac{p}{\left(np\right)} -\frac{p'}{\left(np'\right)} ,\qquad
}
\end{eqnarray}
\begin{equation}\label{ZEqnNum616060}
{\tan \chi_{1}\! =\!\frac{\left(e_{1y} Q_{\!pp'} \right)}{\left(e_{1x} Q_{\!pp'} \right)} \!\equiv \!\tan
\psi _{1},\quad\!\!\!\tan \chi _{2}\! =\! \tan\!\left(\psi _{1} {-}\Delta \right), }
\end{equation}
\begin{eqnarray}\label{ZEqnNum719937}
&&\displaystyle{\alpha _{\pm }\! \left(p,p',\phi _{r} \right)=\alpha _{0\pm }\! \left(p,p'\right)\cdot
g_{1} \!\left(\phi _{r} \right)\cdot g_{2} \!\left(\phi _{r} \right),
}\\
\label{ZEqnNum943439} &&\displaystyle{ \alpha _{0\pm } \!\left(p,p'\right)=\eta _{01}^{} \eta _{02}^{}
\frac{m^{2} }{\omega _{1} \pm \omega _{2} } \left(\frac{1}{\left(np\right)} -\frac{1}{\left(np'\right)}
\right).}
\end{eqnarray}
In Eq. \eqref{ZEqnNum616060} $\psi_{1} =\angle \left({\bf Q}_{pp',\bot },{\bf e}_{1x} \right)$ is the angle
between the projection of the vector ${\bf Q}_{pp'} $ onto the plane $\left(xy\right)$ and the polarization
vector ${\bf e}_{1x} $. In Eqs.~\eqref{ZEqnNum298670} and \eqref{ZEqnNum616060} $\Delta =\angle \left({\bf
e}_{1x} ,{\bf e}_{2x} \right)$  is the angle between vectors of polarization of laser waves. For
definiteness, we assumed that $\omega _{1} >\omega _{2} $.

To obtain the expression for functions $M_{n_{1} n_{2} }^{0} $~\eqref{ZEqnNum504293} and $M_{n_{1} n_{2}
}^{\mu } $~\eqref{ZEqnNum947972} for Figs.~\ref{figure1}(a) and~\ref{figure1}(b), the following replacement
should be performed in Eqs.~\eqref{ZEqnNum159541}-\eqref{ZEqnNum943439}:
\begin{eqnarray}\label{ZEqnNum424834}
(a)\,&&\displaystyle{M_{n_{1} n_{2} }^{0} :\,\, n_{1} =l_{1} {+}s_{1} ,\,\,n_{2} =l_{2} {+}s_{2}
,\,\,p=p_{f} ,\,\,p'=q_{i},\quad
\nonumber}\\
&&\displaystyle{M_{n_{1} n_{2} }^{\mu } :\,\,n_{1} =-s_{1} ,\,\,n_{2} =-s_{2} ,\,\, p=q_{i} ,\,\,p'=p_{i} .
}\end{eqnarray}
\begin{eqnarray}\label{ZEqnNum311016}
(b)\,&&\displaystyle{M_{n_{1} n_{2} }^{0} :\,\,n_{1} =l_{1} {+}s_{1} ,\,\,n_{2} =l_{2} {+}s_{2}
,\,\,p=q_{f}
,\,\,p'=p_{i} ,\quad \nonumber}\\
&&\displaystyle{M_{n_{1} n_{2} }^{\mu } : \,\,n_{1} =-s_{1} ,\,\,n_{2} =-s_{2} ,\,\,p=p_{f} ,\,\,p'=q_{f} .
}
\end{eqnarray}
The replacements $q_{i} \to q_{f} $, $\tilde{\gamma }^{0} \leftrightarrow \hat{\varepsilon }^{*} $ should
be performed in Eqs.~\eqref{ZEqnNum504293} and \eqref{ZEqnNum947972} for Fig.~\ref{figure1}(b).

By virtue of properties of the integer-order Bessel function, its argument determines the characteristic
range of the function order. Thus, arguments $\gamma _{0j} \!\left(p,p'\right)$~\eqref{ZEqnNum680578} and
$\alpha _{0\pm } \!\left(p,p'\right)$~\eqref{ZEqnNum943439} act as multiphoton parameters for ENSB in the
field of two waves. Note that the own specific set of parameters corresponds to the each vertex in
Fig.~\ref{figure1}(a) and \ref{figure1}(b). Notice also that arguments $\gamma _{0j}\! \left(p,p'\right)$
\eqref{ZEqnNum680578}  and $\alpha _{0\pm }\! \left(p,p'\right)$ \eqref{ZEqnNum943439} are essentially
quantum in general case. It is evident that they can have a different order of magnitude with respect to
each other, depending on scattering kinematics.

Arguments $\gamma _{0j} \!\left(p,p'\right)$ \eqref{ZEqnNum680578} are the Bunkin-Fedorov multiphoton
parameters \cite{Bunkin_Fedorov}. They determine the probability of stimulated processes in the field of
each of waves, independently each from other, in Coulomb interaction between particles. Note that the
Bunkin-Fedorov quantum parameter is a multiphoton major parameter in rather broad kinematic range of
scattering angles. This region is called the Bunkin-Fedorov one. The Bunkin-Fedorov parameters may take
values $\gamma _{0j} \sim \eta _{0j} m/\omega _{j} \sim 10^{5} $ for laser fields \eqref{ZEqnNum786952}  in
the general case. However, as can be seen from the expression \eqref{ZEqnNum680578}, these parameters
include the kinematic factor and, accordingly, can have different order of magnitude for the process
different geometry.

Parameters $\alpha _{0\pm } \!\left(p,p'\right)$ \eqref{ZEqnNum943439} are determined by the term that
refers to the interference of the first and second laser waves. They determine the probability of
stimulated correlated processes of absorption or emission of photons of both waves. It is evident from Eq.
\eqref{ZEqnNum943439} that these parameters are determined by the product of the intensities of the first
($\eta _{01} $) and second ($\eta _{02} $) wave, and combination frequencies \cite{Rsp_1994,Rsp_1996}:
\begin{equation} \label{ZEqnNum831260}
\omega _{\pm } \equiv \omega _{1} \pm \omega _{2} .
\end{equation}
Note that if waves' frequency, intensity and polarization are close to each other, then the obtained
amplitude \eqref{ZEqnNum750977}-\eqref{ZEqnNum943439} is transformed into the amplitude for the case of a
single laser wave. The obtained expressions for the amplitude \eqref{ZEqnNum750977}-\eqref{ZEqnNum943439}
in the limit case $\omega _{j} \tau _{j} \to \infty $ coincide also with the corresponding expressions for
the amplitude in the field of two monochromatic waves \cite{Rsp_lys_two,Rsp_lys_two2}.

\section{Kinematic features of ENSB process in two waves}

\subsection{Interference kinematics}

\noindent The value of multiphoton parameters $\gamma _{0j}\! \left(p,p'\right)$~\eqref{ZEqnNum680578} and
$\alpha _{0\pm }\! \left(p,p'\right)$~\eqref{ZEqnNum943439}  for ENSB process in the field of two pulsed
laser waves depends greatly on scattering kinematics. Such a kinematic region (the interference region) can
be distinguished, where quantum parameters $\gamma _{0j}\! \left(p,p'\right)\to 0$  and the parameter
$\alpha _{0\pm }\!\left(p,p'\right)$ becomes the major multiphoton parameter. It was shown in Refs.
\cite{Rsp_1994,Rsp_1996,rsp_leb_en_two,pie_sb,pie_cpp, Rsp_vor_two,Rsp_vor_two2}, that the partial cross
section of the studied process within the interference region can considerably exceed the corresponding
partial cross section in any other geometry, for the case of monochromatic waves. The parametric
interference effect in the problem of nonresonant ENSB was confirmed in Ref. \cite{pie_sb}. The parameters
$\gamma _{0j}\! \left(p,p'\right)$~\eqref{ZEqnNum680578} can be negligible within the interference region:
\begin{equation} \label{ZEqnNum986056}
\gamma _{0j} \left(p,p'\right)\approx 0, \quad  Q_{pp'}^{2} =0.
\end{equation}
Conditions \eqref{ZEqnNum986056} are satisfied when corresponding vectors ${\bf Q}_{pp'} $ are directed
along or against the direction of propagation of both waves ${\bf n}$, i.e. perpendicular to the plane of
polarization $\left({\bf  e}_{jx} ,{\bf  e}_{jy} \right)$. We recall that there are two types of
Bunkin-Fedorov parameters that correspond to  spontaneous emission of a photon by an electron and
scattering of an electron by a nucleus. As it was shown in Refs. \cite{Rsp_lys_two}, kinematics of ENSB
process in the field of two laser waves is identical for the first and second diagram. Moreover,
electron-nucleus scattering and emission of a spontaneous photon occur in the plane, formed by the initial
momentum of an electron and the wave vector of laser field. Therefore, azimuth angles of an electron in the
initial and final states and the azimuth angle of emitted photon are equal:
\begin{equation} \label{ZEqnNum401979}
\varphi '=\varphi _{f} =\varphi _{i} .
\end{equation}
At the same time, the polar angles and momenta of particles are related as:
\begin{equation} \label{ZEqnNum986244}
\cot \frac{\theta '}{2}=a_{i} ,\quad a_{i,f} \equiv \frac{\left|{\bf p}_{i,f} \right|}{\left(np_{i,f}
\right)} \sin \theta _{i,f} ,
\end{equation}
for the angle of emission of a spontaneous photon; and
\begin{equation} \label{ZEqnNum588036}
a_{f} =a_{i} .
\end{equation}
for electron-nucleus scattering. Here, $\theta '{=}\angle \left({\bf n},{\bf k}'\right)$ is the polar angle
of emitted photon; $\theta _{i} {=}\angle \left({\bf n},{\bf p}_{i} \right)$ is the incoming polar angle of
an electron and $\theta _{f}{=}\angle \left({\bf n},{\bf p}_{f} \right)$ is the outgoing polar angle of an
electron. Also, we introduce the angle of scattering of an electron as $\theta {=}\angle \left({\bf p}_{i}
,{\bf p}_{f} \right)$. It is important to note that expressions \eqref{ZEqnNum986244} and
\eqref{ZEqnNum588036} can be fulfilled independently of each other. For example, photon emission occurs
within the interference region (at the angles \eqref{ZEqnNum986244}), and the angle of electron emission
after scattering by a nucleus can be an arbitrary one.

It is easy to obtain that for nonrelativistic electron energies, the expression for the emission angle of a
spontaneously emitted photon \eqref{ZEqnNum986244} simplifies to the form:
\begin{equation} \label{39)}
\theta '=\pi -2v_{i} \sin \theta _{i} ,
\end{equation}
that is, within the interference region the photon is emitted in the opposite direction with respect to the
direction of wave propagation.

It follows from Eq.~\eqref{ZEqnNum482342} and properties of the Bessel function, that $n_{1} =\pm n_{2} $
under the condition \eqref{ZEqnNum986056}. Thus, functions $I_{n_{1} n_{2} }\! \left(\phi \right)$, which
determine the amplitude \eqref{ZEqnNum750977}-\eqref{ZEqnNum159541}, transform into Bessel functions within
the interference region for circular polarization:
\begin{equation} \label{ZEqnNum167805}
I_{n_{1} n_{2} } \left(\phi \right)={\rm e}^{-in_{1} \Delta } J_{n_{1} } \left(\alpha _{\pm }\! \left(\phi
\right)\right)\delta _{n_{1} ,\pm n_{2} } ,
\end{equation}
where, $\delta _{n_{1} ,\pm n_{2} } $ is the Kronecker symbol. We designate the energy, that is absorbed by
an electron from an external laser field under spontaneous emission of a photon, by the way
\begin{equation} \label{ZEqnNum616271}
\omega   \equiv s_{1} \omega _{1} +s_{2} \omega _{2}.
\end{equation}
It is easy to see from Eqs.~\eqref{ZEqnNum298670} and \eqref{ZEqnNum159541}, that within the interference
region \eqref{ZEqnNum401979} and \eqref{ZEqnNum986244}, numbers  of photons of the first and second wave
($s_{1} ,s_{2} $) may differ by a unity. Thus, the quantity \eqref{ZEqnNum616271} can be determined by the
following way:
\begin{equation} \label{ZEqnNum114453}
\omega   =\left\{\begin{array}{l} {s_{1} \left(\omega _{1} +\omega _{2} \right), \qquad \qquad s_{1} =s_{2}
,} \\ {s_{1} \left(\omega _{1} +\omega _{2} \right)+\omega _{2} ,\quad s_{1} =s_{2} -1,} \\ {s_{1}
\left(\omega _{1} +\omega _{2} \right)-\omega _{2} ,\quad s_{1} =s_{2} +1.} \end{array}\right.
\end{equation}
Eq.~\eqref{ZEqnNum114453} indicates, that emission and absorption of photons for the first and second wave
correlate with each other in such a manner. In more detail, the interference kinematics of ENSB process in
a field of two waves is considered in Refs.~\cite{Rsp_lys_two,Rsp_lys_two2}.

\subsection{Resonance conditions}

Along with interference kinematics, in studying of SB of an electron scattered by a nucleus in an external
field, resonance kinematics can be distinguished. Resonance kinematics is related to the possibility of an
electron to fall onto the mass surface in the intermediate state, and is due to the fulfillment of
energy-momentum conservation law for the components of the process of second order in the fine structure
constant \cite{Res_rev_rsp,Res_rev}. Therefore the amplitude of ENSB process in the field of two pulsed
laser waves has the resonant character, when the resonance conditions are met \cite{LebResSB}
\begin{equation} \label{ZEqnNum404382}
q_{i,f}^{2} -m^{2} \lesssim \frac{\left(k_{1} q_{i,f} \right)}{\omega _{1} \tau _{1} } ,
\end{equation}
as it follows from the consideration of Eqs.~\eqref{ZEqnNum122312}, \eqref{ZEqnNum108479} and
\eqref{ZEqnNum537495}. Therefore the four-momentum of an intermediate electron appears near the mass
surface under resonance conditions.

\begin{figure}
\includegraphics[width=6.0cm]{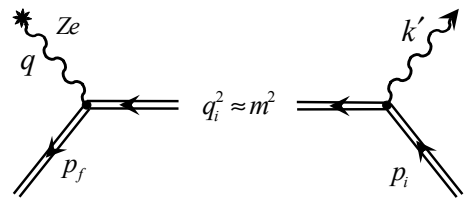}
\caption{\label{fig2}Resonant Feynman diagram for ENSB process in the field of two pulsed laser waves. }
\end{figure}

The interference of resonant amplitudes, which correspond to the diagrams Fig~\ref{figure1}(a) and
Fig~\ref{figure1}(b), can come into existence if conditions \eqref{ZEqnNum404382} are simultaneously
satisfied for an electron in states with four-momenta $q_{i}$ and $q_{f}$. As it was shown in Refs.
\cite{rspNucFiz,rsp_2002,LebResSB}, the described case is realized when an electron is scattered at a small
angle:
\begin{equation} \label{ZEqnNum228670}
\theta \equiv \angle \left( {\bf p}_{i}, {\bf p}_{f} \right)\sim \left(1-{\left(n p_{i}\right)}/{ E_{i}}
\right)\cdot {\omega _{\pm }}/{ \left|{\bf p}_{i} \right|}\ll 1.
\end{equation}
We exclude scattering on given small angles from consideration and will study resonance properties of the
diagram (a) only (see, Fig.~\ref{fig2}).

Let us consider SB of an electron scattered by a nucleus in the field of two pulsed waves in the case when
interference kinematics coincides with resonance kinematics. We fix the angle of emission of a
spontaneously emitted photon in accordance with condition \eqref{ZEqnNum986244}. It is convenient to set
down expressions which determine $q_{i} $ and $q$ \eqref{ZEqnNum249079} for the amplitudes
Fig.~\ref{figure1}(a) as energy-momentum conservation law for each of diagram vertex
\begin{equation} \label{ZEqnNum368785}
\left\{\begin{array}{l} {p_{i} +s_{1} k_{1} +s_{2} k_{2} =q_{i} +k',} \\ {q=p_{f} -q_{i} +\left(l_{1}
+s_{1} \right)k_{1} +\left(l_{2} +s_{2} \right)k_{2} ;} \end{array}\right.
\end{equation}

These laws are fulfilled for only $\omega  >0$ values under the condition \eqref{ZEqnNum404382}. Therefore,
the function $M_{s_{1} ,s_{2} }^{\mu } \left(q_{i} ,p_{i} ,\zeta \right)$ (see, Eq.~\eqref{ZEqnNum947972})
under resonance conditions, determines the amplitude of emission of a photon with the four-momentum $k'$ by
an electron with the four-momentum $p_{i} $ at the expense of the energy $\omega   $ \eqref{ZEqnNum616271}
from external laser waves. The given process in external laser field was considered in Refs. (see,
\cite{Rsp_vor_two,Rsp_vor_two2}). The quantity  $M_{l_{1} +s_{1} ,l_{2} +s_{2} }^{0} \left(p_{f} ,q_{i}
,\zeta \right)$ \eqref{ZEqnNum504293} with respect to the value of the transferred momentum  $q$ (see, the
second quality \eqref{ZEqnNum368785}), determines the amplitude of scattering of an electron with the
four-momentum $q_{i} $ by a nucleus in the field of two pulsed waves with absorption or emission of
$\left|l_{1} +s_{1} \right|$ number of photons of the first wave and $\left|l_{2} +s_{2} \right|$ number of
photons of the second wave. This process for the case of a single electromagnetic wave in the
nonrelativistic limit was studied by Bunkin and Fedorov \cite{Bunkin_Fedorov}. Scattering of an electron by
a nucleus in the field of two waves was studied in Refs.~\cite{rsp_leb_en_two}. Consequently, resonant ENSB
in the field of two pulsed waves is effectively reduced to two consecutive first-order processes with
respect to the fine structure constant.

Taking Eq.~\eqref{ZEqnNum368785} into account, we can find the frequency of the spontaneous photon under
resonance conditions for the range of moderately strong fields \eqref{ZEqnNum786952}. Within the zero order
with respect to the small parameter $\left(\omega_{1} \tau_{1} \right)^{-1} $ the resonant frequency is
specified as \cite{rspNucFiz,LebResSB}:
\begin{equation} \label{ZEqnNum542745}
\omega '_{res} \equiv \omega '_{i}  \frac{1}{1+d_{i} } , \quad  \omega' _{i} =\omega \frac{\left(np_{i}
\right)}{\left(n'p_{i} \right)} ,
\end{equation}
\begin{equation} \label{ZEqnNum474832}
n'=\frac{k'}{\omega '} =\left(1,{\bf n}'\right), \quad  d_{i} =\omega
\frac{\left(nn'\right)}{\left(n'p_{i} \right)} .
\end{equation}

One can see from Eqs. \eqref{ZEqnNum542745} and \eqref{ZEqnNum474832}, that within a rather broad range of
electron energies and scattering angles, we have $d_{i} \ll 1$ (except ultrarelativistic electrons with
energy of order $\sim m^{2} / \omega _{1,2} $, moving within a narrow cone close to the direction of the
momentum of a spontaneous photon). Therefore, resonances are mainly observed when the frequency of a
spontaneous photon is equal to $\omega' _{i} $~\eqref{ZEqnNum542745}.

Within the interference region \eqref{ZEqnNum401979} and \eqref{ZEqnNum986244}, the resonant frequency of a
spontaneously emitted photon can be expressed in terms of the parameters of an initial electron in the
following form:
\begin{equation} \label{ZEqnNum358100}
\omega '_{res} =\omega   \frac{1+a_{i}^{2} }{1-a_{i}^{2} +a_{i}^{} \cot \theta _{i} }
\end{equation}

As can be seen from \eqref{ZEqnNum542745}-\eqref{ZEqnNum358100}, we can separate few characteristic domains
of the frequency $\omega' _{i} $: the nonrelativistic case, $\omega' _{i} \approx \omega _{1,2} $; the
limiting case of ultrarelativistic energies, when an electron moves within a narrow cone related to the
photon from external field $\omega' _{i} \ll \omega _{1,2} $; ultrarelativistic electron moves within a
narrow cone with the spontaneous photon, $\omega' _{i} \gg \omega _{1,2} $; otherwise, $\omega' _{i} \sim
\omega _{1,2} $.

The condition \eqref{ZEqnNum537495} considerably simplifies the integration in the amplitude
\eqref{ZEqnNum122312} and \eqref{ZEqnNum108479}
\begin{eqnarray} \label{ZEqnNum690384}
&&\displaystyle{\int _{-\infty }^{\infty }d\zeta  \frac{\exp \left\{i\zeta \omega _{1} \tau _{1} \left(\phi
_{2} -\phi _{1} \right)\right\}}{q_{i}^{2} -m^{2} +2\zeta \left(k_{1} q_{i} \right)+i0}
\nonumber}\\
&&\displaystyle{ =-i\pi \frac{\exp \left\{i\beta _{i} \left(\phi _{1} -\phi _{2}
\right)\right\}}{\left(k_{1} q_{i} \right)} {\rm H}\left(\phi _{1} -\phi _{2} \right).}
\end{eqnarray}
Here, ${\rm H}\left(\phi _{2} -\phi _{1} \right)$ is the Heaviside function; $\beta _{i} $ is the resonant
parameter that describes resonance ENSB process in the field of two pulsed light waves
\begin{equation} \label{ZEqnNum895805}
\beta _{i} \equiv \frac{q_{i}^{2} -m^{2} }{4\left(nq_{i} \right)} \tau_{1} =\left(1-\frac{\omega '}{\omega
'_{res} } \right)\frac{\omega   \tau_{1} }{2} .
\end{equation}
The expression \eqref{ZEqnNum895805} concludes, that the value of the parameter $\beta _{i} $ is determined
by both process kinematics and external-waves characteristics. The value of this parameter determines how
much the four-momentum of an intermediate electron is close to the mass shell, or how much the frequency of
a spontaneously emitted photon is close to the resonant frequency.

We emphasize, that when the photon is spontaneously emitted within the interference region, the photon
resonant frequency \eqref{ZEqnNum542745} is defined by numbers of photons of the first and second waves.
Thus, observation of the resonant peak on such a frequency in the direction \eqref{ZEqnNum401979} and
\eqref{ZEqnNum986244} can serve as an experimental confirmation of the resonant parametric interference
effect.

\subsection{Multiphoton parameters at specific kinematics}

The amplitude of the considered process is determined by Bessel functions of the integer order. As noted
above, the form of arguments of these functions can vary significantly depending on process kinematics. We
define the explicit form and order of magnitude of the multiphoton parameters $\gamma _{0j}
\left(p,p'\right)$ \eqref{ZEqnNum680578} and $\alpha _{0\pm } \left(p,p'\right)$ \eqref{ZEqnNum943439} for
spontaneous emission of a photon in the interference region \eqref{ZEqnNum401979} and \eqref{ZEqnNum986244}
under resonance conditions \eqref{ZEqnNum542745}.

We denote several useful relationships between convolutions of  four-vectors of particles participating in
ENSB in the case of radiation at a resonant frequency \eqref{ZEqnNum542745}:
\begin{equation} \label{ZEqnNum302804}
\begin{array}{l} {\left(q_{i} p_{i} \right)=m^{2} +\omega   \left(nk'\right),} \\ {\left(k'p_{i} \right)=\omega   \left(nq_{i} \right).} \end{array}
\end{equation}

Hereafter, the upper indices $\left(e\right)$ and $\left(s\right)$ refer to the process of spontaneous
emission of a photon and electron-nucleus scattering, respectively. Bunkin-Fedorov parameters
\eqref{ZEqnNum680578} and the interference parameter \eqref{ZEqnNum943439}, taking relations
\eqref{ZEqnNum302804} into account, can be represented in the form:
\begin{equation} \label{ZEqnNum457340}
\gamma _{0j}^{\left(e\right)} \equiv \gamma _{0j}^{} \left(q_{i} ,p_{i} \right)=2\eta _{0j} \frac{\omega
  }{\omega _{j} } \sqrt{\frac{u'}{u  } \left(1-\frac{u'}{u  } \right)} ,
\end{equation}
\begin{equation} \label{ZEqnNum170044}
\alpha _{0\pm }^{\left(e\right)} \equiv \alpha _{0\pm }^{} \left(q_{i} ,p_{i} \right)=2\eta _{01}^{} \eta
_{02}^{} \frac{\omega   }{\omega _{\pm } } \frac{u'}{u  } .
\end{equation}
Here, relativistic-invariant parameters were introduced:
\begin{equation} \label{57)}
u  =2\omega   \frac{\left(np_{i} \right)}{m^{2} } , \quad   u'_{} =\frac{\left(nk'\right)}{\left(nq_{i}
\right)} .
\end{equation}
Underline that parameters \eqref{ZEqnNum457340} and \eqref{ZEqnNum170044} under resonance conditions become
classical and are actually determined by values of parameters $\eta _{01} $ and $\eta _{02} $. For this
case, within the intensity range \eqref{ZEqnNum786952}, we conclude:
\begin{equation} \label{ZEqnNum860974}
\alpha _{0\pm }^{\left(e\right)}  \sim \eta _{01} \eta _{02} , \quad  \gamma _{0j}^{\left(e\right)}
\lesssim \eta _{0j} .
\end{equation}
Consequently, the partial processes with small numbers of photons ($s_{1,2} =1;0$) give the main
contribution into the resonance cross section within the region of moderately strong fields
\eqref{ZEqnNum786952}, when the Bessel function takes the largest value. It is easy to verify, that for
parallel motion of a spontaneous photon and photons of external pulsed waves ($u'=0 $), resonances are not
observed. At once for the interference region  \eqref{ZEqnNum401979} and \eqref{ZEqnNum986244}, $\gamma
_{0j}^{\left(e\right)} {=}0\to u'{=}u  $ is met.

Bunkin-Fedorov parameters \eqref{ZEqnNum680578} are convenient to be presented through parameters
\eqref{ZEqnNum986244} in the form:
\begin{eqnarray}\label{ZEqnNum147114}
&&\displaystyle{
\gamma _{0j}^{\left(s\right)} \equiv \gamma _{0j}^{} \left(p_{f} ,q_{i} \right)\nonumber}\\
&&\displaystyle{=\eta _{0j} \frac{m}{\omega _{j} } \sqrt{a_{i}^{2} +a_{f}^{2} -2a_{i} a_{f} \cos
\left(\varphi _{f} {-}\varphi _{i} \right)} , }
\end{eqnarray}
Obviously that $\gamma _{0j}^{\left(s\right)} =0$ under the conditions \eqref{ZEqnNum401979} and
\eqref{ZEqnNum588036}. At once the interference parameter \eqref{ZEqnNum943439} does not qualitatively
change and has the following order of magnitude:
\begin{equation} \label{ZEqnNum558081}
\alpha _{0\pm }^{\left(s\right)} \equiv \alpha _{0\pm } \left(p_{f} ,q_{i} \right) \sim \eta _{01} \eta
_{02} \frac{m}{\omega _{\pm } } \gg 1.
\end{equation}
As a result, scattering of an electron by a nucleus in the field of two pulsed waves has a multiphoton
character within the interference region \eqref{ZEqnNum401979} and \eqref{ZEqnNum986244} under resonance
conditions \eqref{ZEqnNum542745}.

\subsection{Amplitude of ENSB process at specific kinematics}

We consider laser-waves circular polarization, when the field strength vectors rotate in the opposite
direction:
\begin{equation} \label{ZEqnNum651359}
\delta _{1} =-\delta _{2} =1.
\end{equation}
We determine the envelope functions for four-potentials of pulsed waves \eqref{ZEqnNum318473} with equal
duration $\tau =\tau _{1} =\tau _{2} $ as Gaussian functions, depending on the dimensionless wave variables
\eqref{ZEqnNum423530}:
\begin{equation} \label{ZEqnNum988180}
g_{1} \left(\phi _{r} \right)=g_{2} \left(\phi _{r} \right)=\exp \left\{-\phi _{r}^{2} \right\}, \quad
r=1,2.
\end{equation}

We consider resonant ENSB process in the field of two pulsed laser waves for the diagram (a) (see,
Fig.~\ref{fig2}), due to absorbtion of a small number of external-field photons is absorbed at the first
vertex:
\begin{equation} \label{ZEqnNum749710}
s_{1,2} =1;0, \quad  s_{1} \omega _{1} +s_{2} \omega _{2} >0, \quad \omega _{1} >\omega _{2} .
\end{equation}
Under such conditions, a spontaneously emitted photon is emitted in a predetermined direction
\eqref{ZEqnNum401979} and \eqref{ZEqnNum986244} with a frequency close to the resonant frequency
\eqref{ZEqnNum542745}.

After uncomplicated computations the amplitude of resonant ENSB in the field of two pulsed waves can
represented in the form:
\begin{equation} \label{ZEqnNum797985}
S_{fi}^{\left(a\right)} =\!\!\sum _{l_{1} ,l_{2} =-\infty }^{\infty }\!\!S_{l_{1} l_{2} }^{\left(a\right)}
,\quad S_{l_{1} l_{2} }^{\left(a\right)} =-i\frac{Ze^{3} \sqrt{\pi } }{\sqrt{2\omega 'E_{i} E_{f} } }
B_{l_{1} l_{2} }^{\left(a\right)} ,
\end{equation}

In view of the estimation \eqref{ZEqnNum860974}, corresponded to spontaneous emission part of the amplitude
\eqref{ZEqnNum947972} can be expanded in a series over small parameters $\eta _{01} $ and $\eta _{02} $.
Then, using Eqs.~\eqref{ZEqnNum690384} and \eqref{ZEqnNum988180}, we integrate over the wave variable $\phi
_{2} $ ($\phi _{1} \to \phi $):
\begin{eqnarray}\label{65)}
&&\displaystyle{B_{l_{1} l_{2} }^{\left(a\right)} =-\frac{i\pi \sqrt{\pi } \tau _{}^{2} }{2\left(nq_{i}
\right)} \int d\phi \,\frac{\exp \left\{iq_{0} \tau \phi \right\}}{q_{\bot }^{2} +\left(q_{0} -q_{z}
\right)^{2} }
\nonumber}\\
&&\displaystyle{ \times  \left(B_{l_{1} l_{2} }^{\left(1,0\right)} +B_{l_{1} l_{2} }^{\left(0,1\right)}
+B_{l_{1} l_{2} }^{\left(1,1\right)} \right).}
\end{eqnarray}
For partial components $B_{l_{1} l_{2} }^{\left(s_{1} ,s_{2} \right)} $, the upper indices correspond to
the number of photons absorbed in the first vertex of the diagram from the first and second waves,
respectively~\eqref{65)}. We underline, that the partial process with $s_{1} =1,\, s_{2} =-1,$ is excluded
from consideration by the choice of the polarization type \eqref{ZEqnNum651359} (the parameter $\alpha
_{0-}^{\left(s\right)} $ does not reveal in this case). Thereby, partial components $B_{l_{1} l_{2}
}^{\left(s_{1} ,s_{2} \right)} $ are determined as:
\begin{equation} \label{ZEqnNum543865}
B_{l_{1} l_{2} }^{\left(1,0\right)} =\exp \left\{2i\beta _{i} \phi {-}\beta _{i}^{2} \right\}\left({\rm
erf}\left(\phi {+}i\beta _{i} \right)+1\right)\tilde{Q}_{l_{1} l_{2} }^{\left(1,0\right)} ,
\end{equation}
\begin{equation} \label{ZEqnNum566462}
B_{l_{1} l_{2} }^{\left(1,1\right)}\! =\!\frac{\sqrt{2} }{2} \exp\! \left\{\!2i\beta _{i} \phi
{-}\frac{\beta _{i}^{2} }{2} \right\}\!\!\left(\!{\rm erf}\!\left(\!\!\sqrt{2} \phi {+}\frac{i\beta _{i}
}{\sqrt{2} } \!\right)\!+1\!\right)\!\tilde{Q}_{l_{1} l_{2} }^{\left(1,1\right)},
\end{equation}
\begin{eqnarray}\label{ZEqnNum910177}
&&\displaystyle{\tilde{Q}_{l_{1} l_{2} }^{\left(s_{1} ,s_{2} \right)} =\bar{u}_{f} M_{l_{1} +s_{1} ,l_{2}
+s_{2} }^{0} \left(p_{f} ,q_{i} ,\phi \right)\left(\hat{q}_{i} +m\right)
\nonumber}\\
&&\displaystyle{\times \left(\varepsilon _{\mu } ^{*} M_{-s_{1} ,-s_{2} }^{\mu } \left(q_{i} ,p_{i}
\right)\right)u_{i} . }
\end{eqnarray}
In Eqs.~\eqref{ZEqnNum543865} and \eqref{ZEqnNum566462},  ``${\rm erf}$'' is the error function. The
function $M_{l_{1} +s_{1} \!,l_{2} +s_{2} }^{0}\! \left(p_{f} ,q_{i} ,\phi \right)$ \eqref{ZEqnNum233516},
\eqref{ZEqnNum298670}-\eqref{ZEqnNum904263} in Eq.~\eqref{ZEqnNum910177}, for field intensities
\eqref{ZEqnNum786952},  is simplified to the form:
\begin{equation} \label{69)}
M_{l_{1} +s_{1} \!,l_{2} +s_{2} }^{0}\! \left(p_{f} ,q_{i} ,\phi \right)=\tilde{\gamma }^{0} I_{l_{1}
+s_{1} \!,l_{2} +s_{2} }\! \left(\chi _{0j}^{\left(s\right)} \!,\gamma _{0j}^{\left(s\right)}\! ,\alpha
_{0+}^{\left(s\right)} ,\phi \right)\!.
\end{equation}

Functions $M_{-s_{1} ,-s_{2} }^{\mu } \left(q_{i} ,p_{i} \right)$ for values $s_{1,2} $
\eqref{ZEqnNum749710}, after decomposition of the special functions \eqref{ZEqnNum482342} in a series over
small parameters $\eta _{01} $ and $\eta _{02} $, assume the form:
\begin{eqnarray}\label{ZEqnNum880035}
&&\displaystyle{M_{-1,0}^{\mu } \left(q_{i} ,p_{i} \right)=-{\rm e}^{i\chi _{1}^{\left(e\right)} }
\frac{\gamma _{01}^{\left(e\right)} }{2} \tilde{\gamma }^{\mu }
\nonumber}\\
&&\displaystyle{+\frac{m\eta _{01} }{4\left(nq_{i} \right)} \hat{\varepsilon }_{1}^{\left(-\right)}
\hat{n}\tilde{\gamma }^{\mu } +\frac{m\eta _{01} }{4\left(np_{i} \right)} \tilde{\gamma }^{\mu }
\hat{n}\hat{\varepsilon }_{1}^{\left(-\right)} , }
\end{eqnarray}
\begin{eqnarray}\label{ZEqnNum506678}
&&\displaystyle{M_{-1,-1}^{\mu } \left(q_{i} ,p_{i} \right)=-{\rm e}^{i\Delta } \left(\frac{\alpha
_{0+}^{\left(e\right)} }{2} -\frac{\gamma _{01}^{\left(e\right)} \gamma _{02}^{\left(e\right)} }{4}
\right)\tilde{\gamma }^{\mu }
}\\
&&\displaystyle{+\frac{m\hat{D}_{-1,-1} \hat{n}\tilde{\gamma }^{\mu }}{4\!\left(nq_{i} \right)}
+\frac{m\tilde{\gamma }^{\mu } \hat{n}\hat{D}_{-1,-1}}{4\!\left(np_{i} \right)}  +\frac{{\rm e}^{i\Delta }
m^{2} \eta _{01} \eta _{02} }{8\!\left(nq_{i} \right)\!\left(np_{i} \right)} \hat{n}n^{\mu } ,\nonumber}
\end{eqnarray}
\begin{equation} \label{72)}
\hat{D}_{-1,-1} =-\frac{1}{2}\! \left(\eta _{01} {\rm e}^{i\chi _{2}^{\left(e\right)} } \gamma
_{02}^{\left(e\right)} \hat{\varepsilon }_{1}^{\left(-\right)} +\eta _{02} {\rm e}^{i\chi
_{1}^{\left(e\right)} } \gamma _{01}^{\left(e\right)} \hat{\varepsilon }_{2}^{\left(-\right)} \!\right)\!.
\end{equation}
Note, that the expression for $B_{l_{1} l_{2} }^{\left(0,1\right)} $ can be easily obtained using $B_{l_{1}
l_{2} }^{\left(1,0\right)} $ \eqref{ZEqnNum543865}, \eqref{ZEqnNum910177}, \eqref{ZEqnNum880035} by
replacement:
$$\eta _{01} \to \eta _{02} , \quad \gamma _{01}^{\left(e\right)} \to \gamma
_{02}^{\left(e\right)} , \quad \chi _{1}^{\left(e\right)} \to \chi _{2}^{\left(e\right)}. $$ Parts of
amplitude are of different orders of the magnitude over parameters $\eta _{01,02} $:
$$M_{l_{1} +s_{1} ,l_{2}
+s_{2} }^{0} \sim 1 , \quad M_{-1,0}^{\mu } \sim \eta _{01} , \quad M_{-1,-1}^{\mu } \sim \eta _{01} \eta
_{02} .$$

Thus, Eqs.~\eqref{ZEqnNum797985}-\eqref{ZEqnNum506678} determine the required resonant amplitude of ENSB
process in the field of two pulsed waves of moderately strong intensities \eqref{ZEqnNum786952} and
circular polarization \eqref{ZEqnNum471197}.

\section{Resonant cross section of ENSB process in two laser waves}

Let us obtain the differential cross section for ENSB process in the field of two pulsed laser
waves\textbf{ }using the resonant amplitudes \eqref{ZEqnNum797985}-\eqref{ZEqnNum506678} by standard
mode~\cite{Landau}
\begin{equation} \label{73)}
d\sigma ^{\left(a\right)} =\frac{\left|S_{f\!i}^{\left(a\right)} \right|^{2} }{v_{i} T} \frac{d^{3} p_{f}
}{\left(2\pi \right)^{3} } \frac{d^{3} k'}{\left(2\pi \right)^{3} } .
\end{equation}
Here, the parameter $T$ is some comparatively great time span. Let us take into account the correlation
$d^{3} p_{f} =\left|{\bf p}_{f} \right|E_{f} dE_{f} d\Omega _{f} $ and $d^{3} k'=\omega '^{2} d\omega
'd\Omega '$. Then, in view the quasi-monochromatic condition \eqref{ZEqnNum305876} the differential cross
section can be presented as a sum of partial components
\begin{equation} \label{ZEqnNum760755}
\frac{d\sigma ^{\left(a\right)} }{d\omega 'd\Omega 'd\Omega _{f} } =\sum _{l_{1} ,l_{2} =-\infty }^{\infty
}\frac{d\sigma _{l_{1} l_{2} }^{\left(a\right)} }{d\omega 'd\Omega 'd\Omega _{f} }  ,
\end{equation}
where $d\sigma _{l_{1} l_{2} }^{\left(a\right)} $ is the partial differential cross section of radiation of
a spontaneous photon into the frequency range $d\omega '$ and the solid angle $d\Omega '$, and the electron
scattering into the solid angle $d\Omega _{f} $, with emission ($l_{1,2} >0$) or absorption ($l_{1,2} <0$)
of $l_{1} $ photons of the first wave and $l_{2} $ photons of the second wave.

Note that the energy conservation law is not strictly realized in the case of a pulsed field. However, in
view of the quasimonochromatic condition \eqref{ZEqnNum305876} the partial cross section $d\sigma _{l_{1}
l_{2} }^{\left(a\right)} $ may be integrated over the final energy of scattered electron $dE_{f} $. It is
easily performed, considering the relationship between the electron energy and the parameter $q_{0} $
\eqref{ZEqnNum249079}. Here we note, that integration could be performed for partial components only.
Therefore, the total cross section \eqref{ZEqnNum760755} is corresponded to the energy spectrum of the
electron in the final state, which is determined by values of the photon numbers $l_{1} $ and $l_{2} $.

Using the amplitudes \eqref{ZEqnNum797985}-\eqref{ZEqnNum506678}, we obtain the differential partial
resonant cross section in the form
\begin{equation} \label{75)}
d\sigma _{l_{1} l_{2} }^{\left(a\right)} =d\sigma _{l_{1} l_{2} }^{\left(1,0\right)} +d\sigma _{l_{1} l_{2}
}^{\left(0,1\right)} +d\sigma _{l_{1} l_{2} }^{\left(1,1\right)} .
\end{equation}
For partial cross sections $d\sigma _{l_{1} l_{2} }^{\left(s_{1} ,s_{2} \right)} $, upper indices
correspond to the number of photons absorbed in the first vertex of diagram (a) from the first ($s_{1} $)
and second ($s_{2} $) waves.
\begin{equation} \label{76)}
\frac{d\sigma _{l_{1} l_{2} }^{\left(s_{1} ,s_{2} \right)} }{d\omega 'd\Omega 'd\Omega _{f} } =\frac{Z^{2}
\alpha r_{e}^{2} m^{2} \omega '\tau ^{2} }{32\pi \left(nq_{i} \right)^{2} q_{}^{4} } P_{l_{1} l_{2}
}^{\left(s_{1} ,s_{2} \right)} \left(\beta _{i} \right)Q^{\left(s_{1} ,s_{2} \right)} .
\end{equation}
Here, the function $P_{l_{1} l_{2} }^{\left(s_{1} ,s_{2} \right)} \left(\beta _{i} \right)$ depends
strongly on the resonant parameter $\beta _{i} $ \eqref{ZEqnNum895805} and, in fact, determines the profile
of the resonance peak ($P_{l_{1} l_{2} }^{\left(1,0\right)} =P_{l_{1} l_{2} }^{\left(0,1\right)} $):
\begin{eqnarray}\label{ZEqnNum267534}
&&\displaystyle{P_{l_{1} l_{2} }^{\left(1,0\right)} \left(\beta _{i} \right)=\exp \left\{-2\beta _{i}^{2}
\right\}
\nonumber}\\
&&\displaystyle{\times \int _{-\rho }^{\rho }\frac{d\phi }{4\rho } \left|{\rm erf}\left(\phi +i\beta _{i}
\right)+1\right|^{2} \left|I_{l_{1} +1,l_{2} } \left(\phi \right)\right|^{2}  , }
\end{eqnarray}
\begin{eqnarray}\label{ZEqnNum295068}
&&\displaystyle{P_{l_{1} l_{2} }^{\left(1,1\right)} \left(\beta _{i} \right)=\exp \{ -\beta _{i}^{2} \}
\nonumber}\\
&&\displaystyle{\times \int _{-\rho }^{\rho }\frac{d\phi }{8\rho } \left|{\rm erf}\!\left(\!\sqrt{2} \phi
+\frac{i\beta _{i} }{\sqrt{2} } \right)+1\right|^{2}\!\! \left|I_{l_{1} +1,l_{2} +1} \left(\phi
\right)\right|^{2} \!.\quad\quad}
\end{eqnarray}
Here, the parameter $\rho =T/\tau $ represents itself the ratio of the observation time  and the
characteristic duration of a laser pulse. This parameter acquires its physical meaning for the concrete
conditions of the process. For example, in the case of external waves in the form of successive laser
pulses, this parameter acquires the physical meaning of the ratio of the distance between neighboring
pulses to the characteristic pulse duration. The function $\left|I_{l_{1} +1,l_{2} } \left(\phi
\right)\right|^{2} $ determines the partial probability of stimulated emission and absorption of
external-waves photons in ENSB in the external field of two waves with the intensity \eqref{ZEqnNum786952}
\cite{L_fun}.

It is easy to consider from expressions \eqref{ZEqnNum267534} and \eqref{ZEqnNum295068} that the resonance
cross section decreases sharply with increasing of the resonance parameter $\beta _{i} $, thus, the cross
section will be of the essence when
\begin{equation} \label{ZEqnNum417584}
\beta _{i} \lesssim 1\Rightarrow \frac{\omega '_{res} -\omega '}{\omega '_{res} } \sim \frac{1}{\omega
  \tau } .
\end{equation}

Summing and averaging over the polarizations of the particles according to general rules \cite{Landau}, it
is easy to obtain an expression for the function $Q^{\left(s_{1} ,s_{2} \right)} $ in the form of a trace
of the product of matrices:
\begin{eqnarray}\label{ZEqnNum342459}
&&\displaystyle{Q^{\left(s_{1} ,s_{2} \right)} =\frac{1}{2} {\rm Sp}\left[\tilde{\gamma }^{0}
\left(\hat{p}_{f} +m\right)\tilde{\gamma }_{0} \left(\hat{q}_{i} +m\right)\right.
\nonumber}\\
&&\displaystyle{\left. \times M_{-s_{1} ,-s_{2} ,\mu }^{} \left(\hat{p}_{i} +m\right) \bar{M}_{-s_{1}
,-s_{2} }^{\mu } \left(\hat{q}_{i} +m\right)\right]. }
\end{eqnarray}
We note, that under resonance conditions \eqref{ZEqnNum404382}, calculation of the trace
\eqref{ZEqnNum342459} is simplified substantially. After uncomplicated computations, we obtain the
differential partial cross section of ENSB in the field of two laser waves \eqref{ZEqnNum318473} and
\eqref{ZEqnNum786952}-\eqref{ZEqnNum471197}, when the photon is emitted within the resonance region
\eqref{ZEqnNum542745} for photon numbers $s_{1,2} $ \eqref{ZEqnNum749710} in the form:
\begin{equation} \label{ZEqnNum156027}
d\sigma _{l_{1} l_{2} }^{\left(s_{1} ,s_{2} \right)} =\frac{\omega '\tau ^{2} E_{i} }{2\left(nq_{i}
\right)^{2} } \cdot dW'_{s_{1} ,s_{2} } \cdot d\sigma _{l_{1} +s_{1} ,l_{2} +s_{2} }^{\left(s\right)} \cdot
P_{l_{1} l_{2} }^{\left(s_{1} ,s_{2} \right)}\! \left(\beta _{i} \right).
\end{equation}

The quantity $d\sigma _{l_{1} +s_{1} ,l_{2} +s_{2} }^{\left(s\right)} $ is the partial differential cross
section of scattering of the electron with the four-momentum $q_{i} $ by a nucleus and transition of the
electron into the final state with the four-momentum $p_{f} $ with emission ($l_{1,2} +s_{1,2} >0$) or
absorption ($l_{1,2} +s_{1,2} <0$) of $\left|l_{1} +s_{1} \right|$ photons of the first wave and
$\left|l_{2} +s_{2} \right|$ photons of the second wave:
\begin{equation} \label{ZEqnNum438289}
\frac{d\sigma _{l_{1} +s_{1} ,l_{2} +s_{2} }^{\left(s\right)} }{d\Omega _{f} } =\frac{2Z^{2} r_{e}^{2}
m^{2} }{{\bf q}^{4} } \left(m^{2} +\left(q_{i} p_{f} \right)+2{\bf q}_{i} {\bf p}_{f} \right),
\end{equation}
where $r_{e} $ is the electron classical radius.

The function $dW'_{s_{1} ,s_{2} } $ is the differential probability per unit time of spontaneous emission
of a photon $k'$ by the initial electron with the four-momentum $p_{i} $ and transition of the electron
into the state with the four-momentum $q_{i} $, at the expense of absorption of $s_{1} $ first-wave photons
and $s_{2} $ second-wave photons. In the general case, this probability can be represented in the form:
\begin{equation} \label{ZEqnNum121598}
\frac{dW'_{s_{1} ,s_{2} } }{d\omega 'd\Omega '} =\frac{\alpha m^{2} }{4\pi E_{i} } W''_{s_{1} ,s_{2} } .
\end{equation}
where the function $W''_{s_{1} ,s_{2} }$ has a rather cumbersome form and coincides with the corresponding
expression for the case of two monochromatic waves in the limit case $\omega _{j} \tau _{j} \to \infty $
\cite{Rsp_vor_two2}.

For partial processes of emission of a spontaneous photon with photon number values $s_{1,2} $
\eqref{ZEqnNum749710}, the general expression can be simplified. Thus, consider
\eqref{ZEqnNum880035}-\eqref{ZEqnNum506678}, we obtain:
\begin{equation} \label{ZEqnNum594446}
W''_{1,0} =\eta _{01}^{2} \left[1+\frac{u'^{2} }{2\left(1+u'\right)} -\frac{4u'}{u  } \left(1-\frac{u'}{u
} \right)\right].
\end{equation}
\begin{eqnarray}\label{ZEqnNum457073}
&&\displaystyle{W''_{1,1} =\eta _{01}^{2} \eta _{02}^{2} \left[-\frac{4u'^{2} }{u ^{2} } u_{\omega} ^{2}
\left(1-\frac{u  u'}{2\left(1+u'\right)} \right)\right.
\nonumber}\\
&&\displaystyle{+\frac{4u'}{u  } u_{\omega}  -\frac{\left(D_{-1,-1} D_{-1,-1}^{*}
\right)}{2} \left(1+\frac{u'^{2} }{2\left(1+u'\right)} \right) }\\
&&\displaystyle{\left. -\frac{4u'^{2} }{mu  } u_{\omega}  \cdot \Re\left\{\left(q_{i} D_{-1,-1}^{*}
\right)-\frac{1}{1+u'} \left(p_{i} D_{-1,-1}^{*} \right)\right\} \right].\nonumber}
\end{eqnarray}
\begin{equation} \label{87)}
u_{\omega}   =1-\frac{\omega  ^{2} }{\omega _{1} \omega _{2} } \left(1-\frac{u'}{u  } \right).
\end{equation}

It is easy to assume, that within the interference region \eqref{ZEqnNum401979} and \eqref{ZEqnNum986244}
($u=u'  $, $u_{\omega}   =1$), for emission of the spontaneous photon, differential probabilities
\eqref{ZEqnNum121598}-\eqref{ZEqnNum457073} are simplified to the form:
\begin{equation} \label{88)}
\frac{dW'_{1,0} }{d\omega 'd\Omega '} =\frac{\alpha m^{2} \eta _{01}^{2} }{4\pi E_{i} }
\left(1+\frac{u'^{2} }{2\left(1+u'\right)} \right),
\end{equation}
\begin{equation} \label{89)}
\frac{dW'_{1,1} }{d\omega 'd\Omega '} =\frac{\alpha m^{2} \eta _{01}^{2} \eta _{02}^{2} }{2\pi E_{i} }
\frac{u'^{2} }{1+u'} .
\end{equation}
Underline, that within the interference region, the probability of spontaneous photon emission at the
expense of a single-photon absorption exceeds greater the probability of absorption of a single photon from
each of waves:
\begin{equation} \label{90)}
\frac{dW'_{1,0} }{dW'_{1,1} } \sim \eta _{01}^{-2} u'^{-2} \sim \frac{m^{2} }{\eta _{01}^{2} \omega
_{1}^{2} } \gg 1.
\end{equation}
Outside the interference region, this ratio is not large: $$dW'_{1,0} /dW'_{1,1} \sim \eta _{02}^{-2} .$$

For electron relativistic energies  ($E \gtrsim m$) and moderately strong intensity of external waves
\eqref{ZEqnNum786952}, the cross section of ENSB \eqref{ZEqnNum760755} and \eqref{ZEqnNum156027} can be
easily summed over partial processes (see, \eqref{ZEqnNum662245}). In this case, the energy corrections
over the field can be neglected, and the cross section \eqref{ZEqnNum438289} transforms to the cross
section of electron scattering by a nucleus in the absence of an external $d\sigma _{l_{1} l_{2}
}^{\left(*\right)} \approx d\sigma _{\rm Mott} $ \cite{Mott}. Finally, we obtain:
\begin{equation} \label{91)}
d\sigma ^{\left(a\right)} =\sum _{s_{1} ,s_{2} }d\sigma _{s_{1} ,s_{2} }  \approx d\sigma _{1,0} +d\sigma
_{0,1} +d\sigma _{1,1} ,
\end{equation}
\begin{equation} \label{ZEqnNum377409}
d\sigma _{s_{1} ,s_{2} } =\frac{E_{i} \tau ^{2} }{2\left(nq_{i} \right)^{2} } \cdot d\sigma _{\rm Mott}
\omega '_{s_{1} ,s_{2} } dW'_{s_{1} ,s_{2} } P_{res}^{\left(s_{1} ,s_{2} \right)} ,
\end{equation}
where the quantities $\omega '_{s_{1} ,s_{2} } $ denote the resonant frequencies  \eqref{ZEqnNum542745} for
the particular values of photon numbers $s_{1,2} $~\eqref{ZEqnNum749710}. Functions $P_{res}^{\left(s_{1}
,s_{2} \right)} \!\left(\beta _{i} \right)$ describe the profile of resonant peaks:
\begin{equation} \label{ZEqnNum154955}
P_{res}^{\left(1,0\right)}\! \left(\beta _{i} \right)=\exp \left\{-2\beta _{i}^{2} \right\}\! \int _{-\rho
}^{\rho }\!\frac{d\phi }{4\rho }\! \left|{\rm erf}\!\left(\phi +i\beta _{i} \right)+1\right|^{2}  ,
\end{equation}
\begin{equation} \label{ZEqnNum615420}
P_{res}^{\left(1,1\right)}\! \left(\beta _{i} \right)=\exp \{ -\beta _{i}^{2} \} \!\int _{-\rho }^{\rho
}\!\frac{d\phi }{8\rho }\! \left|{\rm erf}\!\left(\!\sqrt{2} \phi +\frac{i\beta _{i} }{\sqrt{2} }
\right)+1\right|^{2} \!\! .
\end{equation}

\begin{figure}
\includegraphics[width=7.5cm]{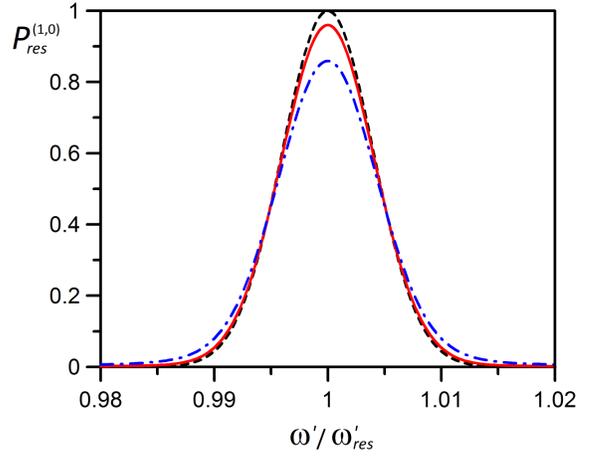}
\caption{\label{figure4}(Color online) The profile of the resonant peak \eqref{ZEqnNum154955} in the cross
section of ENSB in two pulsed laser waves. Electron parameters in the initial state: $E_{i} =1.02$ {\rm
MeV}, $\theta _{i} =163^{\circ } $. External field parameters: $\eta _{01} =\eta _{02} =0.1$,  $\omega _{1}
=2.35$~{\rm eV}, $\omega _{2} =1$~{\rm eV}, $\tau =0.1$~{\rm ps}. The solid red curve corresponds to the
value $\rho =5$, dash-dotted blue curve corresponds to the value $\rho =\sqrt{2} $. The dashed curve shows
the Gaussian function $\exp \left\{-2\beta _{i}^{2} \right\}$. }
\end{figure}

Figure~\ref{figure4} presents the function $P_{res}^{\left(1 ,0 \right)}$ \eqref{ZEqnNum154955} as
dependence on the frequency of spontaneously emitted photon (see, Eq.~\eqref{ZEqnNum895805}). Obliviously
that the substantial range of values of the energy of spontaneously emitted photon is sufficiently narrow
and is determined by the condition \eqref{ZEqnNum417584}. At once the decreasing of a function value has an
exponential character.

We distinguish the quantity $\omega '_{e} $ as value of the spontaneous-photon frequency, for which the
value of the peak-profile function falls in $\exp$ times ($\beta _{ie} =1/\sqrt{2}$). And determine the
transit width of the resonance in ENSB as the difference between the resonant frequency of the
spontaneously emitted photon and $\omega '_{e} $:
\begin{equation} \label{ZEqnNum827040}
\beta _{ie} =1/\sqrt{2}\Rightarrow \Gamma _{\omega '} \equiv \omega '_{res} -\omega '_{e}
=\frac{\sqrt{2}\omega '_{res} }{\omega   \tau } .
\end{equation}
It is characteristically, that the transit width of resonance $\Gamma _{\omega '} $  \eqref{ZEqnNum827040}
is inversely proportional to the pulse duration. The resonance transit width can also be entered by
representation of the resonant peak profile in the form of the Lorentz function~\cite{LebResSB}. The value
for the transit width in these two cases differs insignificantly. It is easy to demonstrate that in our
consideration the transit width is much greater than radiation width of a resonance~\cite{LebResSB}.

The functions $P_{res}^{\left(s_{1} ,s_{2} \right)} \!\left(\beta _{i} \right)$
\eqref{ZEqnNum154955}-\eqref{ZEqnNum615420} in substance describes the shape of the bremsstrahlung spectrum
for ENSB within the resonance region. Note that outside the resonance region the bremsstrahlung spectrum is
determined by nonresonant properties of the SB cross section; this cross section for moderately strong
fields in the low-frequency region coincides with the cross section of ENSB process in the absence of an
external field \cite{Nonres_sb}.

The differential cross section \eqref{ZEqnNum377409} can be integrated within the resonance region over the
energy of the spontaneously emitted photon $\omega '$. The smallness of the transit width of the resonance
conduces the fact that the dependence on the energy $\omega '$ should be retained only in the resonance
parameter $\beta _{i} $. Consequently, using the relation $d\omega '=-d\beta _{i} \cdot \sqrt{2}\Gamma
_{\omega '} $, we obtain the resonant cross section of ENSB in the two-waves field for the spontaneous
emission of a photon within the interference region \eqref{ZEqnNum401979} and \eqref{ZEqnNum986244} in the
form:
\begin{eqnarray}\label{ZEqnNum826701}
&&\displaystyle{\frac{d\sigma ^{\left(a\right)} }{d\Omega 'd\Omega _{f} } =\sqrt{\frac{\pi }{2} }
\frac{E_{i} \tau }{\left(nq_{i} \right)^{2} \omega   }
\nonumber}\\
&&\displaystyle{\times \frac{d\sigma _{\rm Mott} }{d\Omega _{f} } \left({\omega '}_{\!1,0}^{2}
\frac{dW'_{1,0} }{d\Omega '} + {\omega' }_{\!0,1}^{2} \frac{dW'_{0,1} }{d\Omega '} \right) .}
\end{eqnarray}
Obtained resonance cross section \eqref{ZEqnNum826701} is valid for intensities of moderately strong
fields~\eqref{ZEqnNum786952}, for electron scattering at large angles $\theta \gg {\omega / \left|{\bf
p}_{i} \right|} $ \eqref{ZEqnNum228670}.

Note that the cross section of SB in the absence of an external field $d\sigma _{\rm BH} $ (the
Bethe-Heitler cross section \cite{bethe34}) in the considered case can be factorized into the product of
the cross section for the elastic scattering of an electron by a nucleus $d\sigma _{\rm Mott} $ \cite{Mott}
and the probability of emission of a photon $dW_{k' } $ \cite{Landau}
\begin{equation} \label{ZEqnNum720280}
\frac{d\sigma _{\rm BH} }{d\Omega 'd\Omega _{f} } =\frac{d\sigma _{\rm Mott} }{d\Omega _{f} } \cdot
\frac{dW_{k' } }{d\Omega '} ,
\end{equation}
\begin{equation} \label{98)}
\frac{d\sigma _{\rm Mott} }{d\Omega _{f} } =\frac{2Z^{2} r_{e}^{2} m^{2} }{{\bf q}^{4} } \left(E_{i} E_{f}
+m^{2} +{\bf p}_{i} {\bf p}_{f} \right),
\end{equation}
\begin{equation} \label{ZEqnNum832902}
\frac{dW_{k' } }{d\Omega '} =\frac{\alpha }{4\pi ^{2} } \cdot \left\{{\bf q}^{2} -\left({\bf n'} {\bf
q}\right)^{2} \cdot \frac{m^{2} }{\kappa '_{i} \kappa '_{f} } \right\}\cdot \frac{d\omega '}{\omega '\kappa
'_{i} \kappa '_{f} } .
\end{equation}
\[{\bf q}= {\bf p}_{f} -{\bf p}_{i},\quad \!\!\kappa _{i,f} =E_{i,f} -{\bf n}{\bf p}_{i,f} ,\quad\!\!
\kappa '_{i,f} =E_{i,f} -{\bf n'}{\bf p}_{i,f} \]

In further analysis, we consider the ratio of the resonance cross section of ENSB in the field of two
waves \eqref{ZEqnNum826701} for emission of a spontaneous photon within the interference region
\eqref{ZEqnNum401979} and \eqref{ZEqnNum986244} to the bremsstrahlung cross section in the absence of an
external field \eqref{ZEqnNum720280} - \eqref{ZEqnNum832902}
\begin{equation} \label{100)} R_{{\rm res}}
=R_{1,0} +R_{0,1} ,
\end{equation}
\begin{equation} \label{ZEqnNum403439}
R_{1,0} =\frac{d\sigma _{res}^{\left(1,0\right)} }{d\sigma _{\rm BH} } =\frac{\pi \sqrt{2\pi } }{8} \eta
_{01}^{2} \left(\omega _{1} \tau \right)^{2} \frac{\omega '_{res} }{\omega _{1} } \frac{m^{2} }{{\bf
p}_{i}^{2} } \cdot f_{1,0} .
\end{equation}
Here, the function $f_{1,0} \sim 1$ and has the form
\begin{equation} \label{102)}
f_{1,0} =\frac{\kappa '_{\!f} /\kappa '_{i} }{4\sin ^{2} \!\left(\theta /2\right)-\left(\cos \theta '_{\!f}
-\cos \theta '_{i} \right)^{2} m^{2} /\kappa '_{i} \kappa '_{\!f} } ,
\end{equation}
where $\theta =\angle \left( {\bf p}_{i}, {\bf p}_{f} \right) $ is the electron scattering angle.

\begin{figure}
\includegraphics[width=8.0cm]{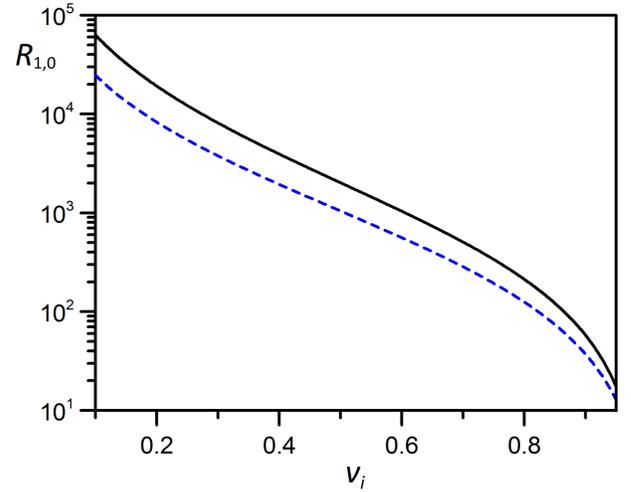}
\caption{\label{R}(Color online) The ratio of the resonance cross section of ENSB in the field of two
pulsed waves to the bremsstrahlung cross section in the absence of an external field \eqref{ZEqnNum403439}
as a function of the initial electron velocity at the fixed ingoing angle $\theta _{i} =163^{\circ } $.
External field parameters: $\eta _{01} =\eta _{02} =0.1$,  $\omega _{1} =2.35$ {\rm eV}, $\omega _{2} =1$
{\rm eV}, $\tau =0.1$ {\rm ps}. The solid black curve corresponds to the value of outgoing electron angle
$\theta _{f} =10^{\circ } $, dashed blue curve corresponds to $\theta _{f} =30^{\circ } $.}
\end{figure}

The dependence of the function $R_{1,0}$~\eqref{ZEqnNum403439} on the electron initial velocity is
presented in Figure~\ref{R}. From Figure~\ref{R} we conclude, that the resonant differential cross section
of ENSB process with simultaneous registration of both emission angles of the spontaneous photon and the
scattered electron, can exceed by 4-5 orders of magnitude the corresponding cross section in the absence of
an external field. The highest value (5 orders) of this ratio is in the case of nonrelativistic electron
energies. On the contrary the ratio \eqref{ZEqnNum403439} decreases sharply for ultrarelativistic electron
energies.

As noted above, for bremsstrahlung of an electron elastically scattered by a nucleus in an external field,
the final energy of electron is described a certain spectrum even for a fixed value of the energy of the
spontaneous emitted photon. The value of the electron energy depends on the number of external-field
photons of forcedly emitted or absorbed by an electron. The distribution over the energy is determined by
the probability of partial processes of stimulated emission and absorption. As it was shown in the
Refs.~\cite{rsp_leb_en_two,pie_cpp}, the parametric interference effect, associated with correlated
emission and absorption of photons of the first and second laser waves, manifests itself in a qualitative
change of the form of the spectrum of final particles within the interference region. This indicates that
stimulated emission and absorption by an electron within the interference region is correlated. Estimates
show that the resonant cross section of ENSB in the field of two pulsed laser waves within the interference
region in two order of magnitude may exceed corresponding cross section in the Bunkin-Fedorov kinematic
region.

The obtained results may be experimentally verified, for example, by scientific facilities at sources of
pulsed laser radiation (SLAC, FAIR, XFEL, ELI, XCELS).

\section{Conclusions}

Performed study of resonant ENSB in the field of two pulsed laser waves results in following conclusion:

Spontaneous bremsstrahlung of an electron scattered by a nucleus in the external laser field of two pulsed
waves is characterized by the presence of a specific kinematic region. Within this region, the process
cross section has the resonant character, and stimulated emission and absorption of photons of the first
and second waves proceed in a correlated manner. Under resonance conditions, the considered second-order
process with respect to the fine-structure constant effectively decomposes into two first-order
processes.

 The resonant differential cross
section of ENSB process with simultaneous registration of both emission angles of the spontaneous photon
and the scattered electron, can exceed by 4-5 orders of magnitude the corresponding cross section in the
absence of an external field. The highest value (5 orders) of this ratio is in the case of nonrelativistic
electron energies. On the contrary it decreases sharply for ultrarelativistic electron energies.

 The correspondence between the emission angle and the final-electron
energy is established in the kinematic region where the resonant parametric interference effect is
manifested. The resonant cross section of ENSB in the field of two pulsed laser waves within the
interference region in two order of magnitude may exceed corresponding cross section in the Bunkin-Fedorov
kinematic region for nonrelativistic electron energy.

\nocite{*}

\bibliography{sb_pie} 

\end{document}